# Magnetocaloric effect and its implementation in critical behavior study of $Mn_4FeGe_{3-x}Si_x$ intermetallic compounds

Madhumita Halder and S. M. Yusuf [a)]

*Solid State Physics Division, Bhabha Atomic Research Centre, Mumbai 400085, India*

A. K. Nigam

*Tata Institute of Fundamental Research, Homi Bhabha Road, Mumbai 400005, India*

## Abstract

Magnetocaloric effect in $Mn_4FeGe_{3-x}Si_x$ compounds has been studied by dc magnetization measurements. For the parent compound $Mn_4FeGe_3$, the paramagnetic to ferromagnetic transition temperature $T_C$ is above room temperature (320 K), which initially remains constant for small Si substitution at the Ge site and then decreases marginally with an increase in Si concentration. A large change in magnetic entropy at the $T_{C,}$ under a magnetic field variation of 50 kOe, with typical values of 5.9, 6.5, 5.9 and 4.4 J kg$^{-1}$ K$^{-1}$ for $x$ = 0, 0.2, 0.6, and 1 samples, respectively, along with a broad operating temperature range and a negligible hysteresis make $Mn_4FeGe_{3-x}Si_x$ series a promising candidate for magnetic refrigerant material around room temperature. $Mn_4FeGe_{3-x}Si_x$ series is found to undergo a second-order magnetic phase transition. The field dependence of the magnetic entropy change has been brought out and implemented it to deduce the critical exponents. The critical behavior study shows that the magnetic interactions for $x$ = 0 and 0.2 samples have two different behaviors below and above $T_C$. Below $T_C$, it follows the mean field theory with long-range magnetic interaction and above $T_C$ it follows the Heisenberg three-dimensional model with short-range or local magnetic interaction. The magnetic exchange interactions for the $x$ = 0.6 and 1 samples follow the mean-field theory.

[a)] Electronic mail: smyusuf@barc.gov.in FAX: +91 22 25505151

## I. INTRODUCTION

Magnetocaloric effect (MCE) is a topic of current interest because of its possible application in magnetic refrigeration[1, 2] near room temperatures. Magnetic refrigeration is an environmental friendly cooling technology with a better cooling efficiency. It can replace the existing vapor-compression refrigeration.[3, 4] The current research in literature aims in probing new materials showing a giant magnetocaloric effect (GMCE) with a small change in magnetic field near room temperature. For example, the rare earth element gadolinium (Gd) has been widely investigated for its use as an active magnetic refrigerant near room temperature.[5] Later it was shown that $Gd_5Si_2Ge_2$ [5]alloy exhibits GMCE with a typical value of 18.5 J kg$^{-1}$ K$^{-1}$. Some other compounds which show a large MCE, are $MnFe(P_{1-x}As_x)$,[6] $La(Fe_{13-x}Si_x)$,[7] Ni-Mn-Ga Heusler alloys,[8, 9] perovskites *etc.*[10, 11] However, most of these materials undergo a first-order magnetic phase transition. The change in magnetic entropy is large in such type of materials, but they exhibit large thermal and field hystereses on variation of magnetization with temperature and magnetic field, respectively, which act as a limitation for their usage in practical applications. Hence, there is a need for searching new advanced magnetic materials with a second-order magnetic phase transition, showing a large reversible magnetic entropy change $\Delta S_M$ at low applied fields. For this, it is important to know the field dependence of $\Delta S_M$ of a given magnetic refrigerant material. Recently, there has been some studies[12, 13] based on the phenomenological universal curve for the field dependence of $\Delta S_M$ as proposed by Franco *et al.*[14] Moreover, recently Fan *et al* [15] showed that this universal curve can be used to evaluate the critical exponents and to study the nature of magnetic phase transition in materials.

Generally, rare-earth based intermetallic compounds are investigated for MCE as they show a large entropy change near their magnetic transition temperatures. Among transition metal based materials, Heusler alloys are the most studied. Among other transition metal based intermetallic compounds (non Heusler alloys), $Mn_5Ge_3$ is a ferromagnet with the magnetic transition temperature near room temperature (296 K).[16] $Mn_5Ge_3$ is a multifunctional compound showing a large MCE[17] and a spin polarization.[18, 19] By replacing one Mn atom in $Mn_5Ge_3$ with Fe, the transition temperature and the spin polarization were enhanced.[20] $Mn_4FeGe_3$ has $D8_8$ crystal structure, same as that of $Mn_5Ge_3$. In the present study, we have investigated the MCE of $Mn_4FeGe_3$ intermetallic compound as well as for the Si substituted compounds of $Mn_4FeGe_{3-x}Si_x$ series by dc magnetization. We have also investigated the magnetic phase transition in detail by studying the field dependence of $\Delta S_{M,}$ and implemented it to deduce the critical exponents. The obtained values of the critical exponents obey the scaling theory, indicating that the values thus obtained are reliable.

## II. EXPERIMENTAL DETAILS

The polycrystalline $Mn_4FeGe_{3-x}Si_x$ samples ($x$ = 0, 0.2, 0.6 and 1) were prepared by an arc-melting method under an argon atmosphere, with the constituent elements Mn (99.99% purity), Fe (99.99% purity), Ge (99.9999% purity) and Si (99.9995% purity). For a better chemical homogeneity, the samples were re-melted many times. The powder x-ray diffraction (XRD), using the Cu-$K_\alpha$ radiation in the 2θ range of 10 – 90º with a step of 0.02º, was carried out on all samples at room temperature. The dc magnetization measurements were carried out on all samples using a Physical Property Measurement System (PPMS, Quantum Design) as a function of temperature and magnetic field. The field-cooled (FC) magnetization measurements were carried out in the warming cycle over the temperature range of 5–350 K in the presence of 200 Oe magnetic field. Magnetization as a function of magnetic field was measured for all samples at 5 K over a field variation of ± 50 kOe. The magnetization isotherms for all samples were recorded at various temperatures with an interval of 5 K up to a maximum applied field of 50 kOe.

## III. RESULTS AND DISCUSSION

XRD patterns [Fig.1 (a)] at room temperature confirm that the samples were in single phase with $D8_8$ hexagonal $Mn_5Si_3$-type crystal structure and space group $P6_3/mcm$. Mn atoms occupy two different crystallographic sites i.e. 4*d* and 6*g*, while Ge/Si atoms occupy the 6*g* site. Fe atom goes to the 4*d* site. The values of lattice constants for the $x = 0$ sample are close to those reported by Chen *et al.*[20] The lattice constants decrease with an increase in the Si content [Fig.1 (b)], since Si is a smaller atom than Ge. The $c/a$ ratio remains almost constant indicating a uniform contraction of the unit cell volume.

Figure 2 (a) shows the FC magnetization ($M$) versus temperature ($T$) curves under an applied field ($H$) of 200 Oe for all samples. The paramagnetic to ferromagnetic transition temperature (the Curie temperature $T_C$) is derived from the minima of the d$M$/d$T$ vs $T$ curves. $T_C$ initially remains constant for small value of $x$ and then decreases marginally with the increase in Si substitution (320, 320, 319 and 318 K for $x$ = 0, 0.2, 0.6 and 1, respectively). The decrease in $T_C$ could be due to the decrease in Mn-Mn interaction which occurs due to a reduction in the Mn-Mn distance.[21] Similar results were reported for the $Mn_5Ge_{3-x}Si_x$ series.[22] The transition also broadens on Si substitution. Figure 2 (b) shows the $M$ vs $H$ curves at 5 K over a field range of ± 50 kOe (covering all four quadrants) for all four samples. Here, the field hysteresis is negligible indicating a second order magnetic phase transition. The saturation magnetization at 5 K remains almost constant for all samples with the ordered magnetic moments of 2.41, 2.48, 2.44 and 2.47 $\mu_B$/ metal atom for $x$ = 0, 0.2, 0.6 and 1 samples, respectively. Chen *et al* [20] reported that the ordered magnetic moment for the parent compound was to be 2.35 $\mu_B$/ metal atom which is close to that observed by us. It is known that the substitution of Si or Ga at the Ge site in $Mn_5Ge_3$ reduces the ordered magnetic moment but this reduction is marginal for a small substitution.[17, 22, 23] In the present study, due to a small substitution, the change in the ordered magnetic moment is not observable. Probably, this could be due to the mixing of the 3*d* electronic states of the Mn atoms with the 3*p* electronic states of the Si atoms.

Figure 3 shows a series of magnetization isotherms measured at various temperatures for the samples with $x$ = 0, 0.2, 0.6, and 1. The magnetic entropy change $\Delta S_M$ was calculated from the magnetization isotherms as follows[24]

$$\Delta S_M(H,T) = \mu_0 \int_0^H \left( \frac{\partial M(H,T)}{\partial T} \right)_H dH \qquad (1)$$

with $\mu_0$ being the permeability in vacuum. For magnetization measurements made at discrete intervals of temperature and field, $\Delta S_M$ $(H, T)$ can be approximated as

$$\Delta S_M(H,T) = \mu_0 \sum_i \frac{M_{i+1}(T_{i+1},H) - M_i(T_i,H)}{T_{i+1} - T_i} \Delta H \qquad (2)$$

Equation 2 is used to estimate the values of $\Delta S_M$. Figure 4 shows the variation of -$\Delta S_M$ with temperature for all four samples. The maximum value of $\Delta S_M$ is found to be around $T_C$ and it increases with the increase in the applied magnetic field. At a field variation of 50 kOe, -$\Delta S_M$ values are found to be 5.9, 6.5, 5.9 and 4.4 J $kg^{-1}$ $K^{-1}$ for $x$ = 0, 0.2, 0.6, and 1 samples, respectively. We also observe that at a lower substitution of Si at the Ge site (i.e from $x$ = 0 to 0.2), the $T_C$ almost remains constant but the -$\Delta S_M$ value increases. With further increase in the Si substitution, the value of -$\Delta S_M$ decreases. The values of -$\Delta S_M$ for these samples are quite appreciable. However, the present values are significantly smaller than those of other GMCE materials like $Gd_5Ge_2Si_2$ (18.5 J $kg^{-1}$ $K^{-1}$)[5] and $MnFeP_{0.45}As_{0.55}$ (18 J $kg^{-1}$ $K^{-1}$).[6] The width of the -$\Delta S_M$ curves increases on Si substitution. For $x$ = 0.6 and 1 samples, the value of -$\Delta S_M$ almost remains constant over a wide temperature range, and then decreases gradually as the temperature shifts away from $T_C$ (Fig. 4). This is because the magnetic transition spreads over a broad temperature range. This could be due to the disorder at the Ge site. Thus by substituting Si at the Ge site, the operating temperature range increases, which is important for practical applications.[25] Another useful parameter which decides the efficiency of a magnetocaloric material is the relative cooling power (RCP) or the refrigerant capacity. The RCP has been calculated by a method suggested in the literature.[14, 26] The calculated values of RCP under a magnetic field variation of 50 kOe, are 244, 268, 295 and 431 J $kg^{-1}$ for $x = 0$, 0.2, 0.6 and 1 samples, respectively. Thus, a large value of $-\Delta S_M$ with a broad operating temperature range and a negligible hysteresis, makes $Mn_4FeGe_{3-x}Si_x$ series, a potential candidate for a magnetic refrigerant material around room temperature.

We have used the field dependence of $\Delta S_M$ to investigate the critical behavior for all samples of the $Mn_4FeGe_{3-x}Si_x$ series. According to the scaling hypothesis,[27] a second-order magnetic phase transition near $T_C$ is characterized by a set of critical exponents β, γ, and δ. The conventional method to calculate the critical exponents is the modified Arrott plots method, based on the Arrott-Noakes equation of state.[28] In this method, one has to make an initial choice of the critical exponents which is difficult, and affects the final value. Due to the drawback of this method, we have used the field dependence on magnetic entropy change to deduce the critical exponents as suggested by Fan *et al.*[15] The field dependence of magnetic entropy has been used earlier by many authors in literature including us.[12, 14, 29] Figure 5 depicts the Arrott plot ($M^2$ vs $H/M$) for the $x = 0$, 0.2, 0.6 and 1 samples. According to the Banerjee criteria,[30] a positive slope of the $M^2$ vs $H/M$ curves corresponds to the second-order phase transition, while a negative slope corresponds to a first-order phase transition. The observed positive slope of the $M^2$ vs $H/M$ curves indicates that the paramagnetic to ferromagnetic transition is of second-order in nature for the parent sample as well as for all Si substituted samples. According to the scaling hypothesis,[27] at $T_C$, the exponent δ relates $M$ and $H$ by

$$M\left(T_C\right) = DH^{1/\delta} \qquad (3)$$

where $D$ is the critical amplitude. To find the value of δ, the $M$ ($T_C$) versus $H$ isotherm is plotted on the log-log scale [ Fig. 6 (a)] for the $x$ = 0, 0.2, 0.6, and 1 samples. According to Eq. (3), this should be a straight line with a slope 1/δ. From the linear fit of the straight line, the obtained values of δ are 4.20(4), 4.27(3), 2.75(1) and 2.41(6) for $x$ = 0, 0.2, 0.6 and 1 samples, respectively. The critical exponents β, γ, and δ are related to each other by the Widom scaling relation: $\delta = 1 + \gamma / \beta$.[31] The field dependence of $\Delta S_M$ given by the following equation:[14]

$$\Delta S_M \big|_{T=T_C} \propto H^n \qquad \text{where } n = 1+1/\delta(1-1/\beta) \qquad (4)$$

Using the Widom scaling relation and Eq. (4), we obtain the values of exponents β and γ. The peak of the $\Delta S_M$ versus $H$ curve has been depicted in Fig. 6 (b). The values of $n$, obtained from the fitting using Eq. (4), are 0.72(1), 0.71(2), 0.80(3) and 0.86(1) for $x$ = 0, 0.2, 0.6 and 1 samples, respectively. The values of the critical exponents β and γ, thus obtained are 0.459 and 1.47 for the $x$ = 0 sample, 0.445 and 1.457 for the $x$ = 0.2 sample, 0.645 and 1.127 for the $x$ = 0.6 sample, and 0.748 and 1.054 for the $x$ = 1 sample, respectively. The values of the critical exponents for the $x$ = 0 and 0.2 samples are close to that obtained for $Mn_5Ge_3$.[32] The derived values of the critical exponents lie between three dimensional Heisenberg model[33] and mean-field theory.[27] Here, it is interesting to note that for the $x$ = 0 and 0.2 samples, none of the theoretical models i.e., the three dimensional Heisenberg or the Ising model with short-range magnetic exchange interaction and, the mean-field theory with a long-range magnetic exchange interaction describes the magnetic phase transition completely. Magnetic interactions in the present $x$ = 0 and 0.2 samples show two different behaviors above and below $T_C$. Below $T_C$, the value of critical exponent β is close to that obtained from the mean field theory (β = 0.5)[27] indicating a long-range magnetic interaction. The value of the critical exponent γ is close to that obtained from the Heisenberg three-dimensional model (γ =

1.396)[33] indicating that a short-range or a local magnetic interaction is present above $T_C$. Similar phenomenon has also been observed in other compounds.[34, 35] The derived values of the critical exponents (both β and γ) for the $x$ = 0.6 and 1 samples are close to that of the mean-field theory, though the value of β is slightly higher than that predicted from the mean-field theory. This suggests that with a further higher substitution of Si, the system follows the mean-field theory with a long-range magnetic interaction both below and above $T_C$.[36] The value of β increases with the increase in Si substitution (i.e. for the $x$ = 0.6 and 1 samples), indicating a slower growth of the ordered moment with decreasing temperature. This is evident from the broadening of the $M$ vs $T$ and $-\Delta S_M$ vs $T$ curves, indicating a slow magnetic phase transition. The slower growth of the ordered moment with the decreasing temperature results in a decrease in the value of $-\Delta S_M$ (for the $x$ = 1 sample) as the change in magnetization with temperature is not large in that case.

According to the scaling hypothesis,[37] $M(H,\varepsilon)$ [ε is the reduced temperature $(T-T_C)/T_C$] is a universal function of $T$ and $H$, and the experimental $M(H,\varepsilon)$ curves are expected to collapse into the universal curve

$$M(H,\varepsilon) = \varepsilon^{\beta} f_{\pm}(H/\varepsilon^{\beta+\gamma}) \tag{5}$$

with two branches, one for temperatures above $T_C$ and the other for temperatures below $T_C$. Here $f_+$ for $T > T_C$ and $f_-$ for $T < T_C$ are regular functions. The critical exponent analysis can be justified by the $M\varepsilon^{-\beta}$ vs $H\varepsilon^{-(\beta+\gamma)}$ plot. According to Eq. (5) all data should fall on one of the two curves. The scaled data for all samples are plotted on a log scale as shown in Fig. 7. It can be seen that all data fall on either of the two branches of the universal curve, one for temperatures above $T_C$ and the other for temperatures below $T_C$. This indicates that the critical exponents obtained from field dependence of $\Delta S_M$ are reasonably accurate.

The field dependence of the magnetic entropy change curve helps us to predict the response of a particular material under different experimental conditions which can be useful for designing new materials for magnetic refrigeration. Thus, a study of the MCE for a particular material is not only important from its practical application point of view but it also provides a tool to understand the properties of the material. In particular, the details of the magnetic phase transition and critical behavior of a given material can be obtained by studying the MCE of the material. It is also useful for studying similar materials such as alloy series with small compositional changes since the critical exponents do not change much in that case.

## IV. SUMMARY AND CONCLUSIONS

Here, we have investigated the MCE for the parent compound $Mn_4FeGe_3$ as well as for the Si substituted compounds of the $Mn_4FeGe_{3-x}Si_x$ series. We observe that on substituting Si at the Ge site in $Mn_4FeGe_{3-x}Si_x$, $T_C$ initially remains constant for small value of $x$ and then decreases marginally with the increase in Si substitution. The decrease in $T_C$ could be due to the decrease in Mn-Mn interaction which occurs due to a reduction in the Mn-Mn distance. For these samples, the values of $-\Delta S_M$ are comparable with that of other GMCE materials near room temperature. The substitution of Si at the Ge site causes an increase in the operating temperature range for MCE. Thus, a large value of $-\Delta S_M$ with a broad operating temperature range and a negligible hysteresis, makes $Mn_4FeGe_{3-x}Si_x$ series, a potential candidate for a magnetic refrigerant material around room temperature. We have studied the field dependence of $\Delta S_M$, and implemented it to investigate the critical behavior for the all samples of $Mn_4FeGe_{3-x}Si_x$ series. The field and temperature dependent magnetization behavior follows the scaling theory, and all data points fall on the two distinct branches, one for $T < T_C$ and the other for $T > T_C$ indicating that the critical exponents thus obtained are reasonably accurate. The magnetic interactions in the present $x = 0$ and 0.2 samples show two different behaviors above and below $T_C$. Below $T_C$, it follows the mean field theory with a long-range magnetic interaction and above $T_C$ it follows the Heisenberg three-dimensional model with a short-range or a local magnetic interaction. The magnetic exchange interactions for the $x = 0.6$ and 1 samples follow the mean-field theory.

## ACKNOWLEDGMENTS

M. H. acknowledges the help provided by D. D. Buddhikot for the magnetization measurements and, A. K. Rajarajan and M. D. Mukadam for the sample preparation. M. H. also thanks Homi Bhabha National Institute, Department of Atomic Energy, India, for the fellowship.

**List of Figures**

FIG. 1: (Color online) (a) x-ray diffraction patterns for the $x = 0$ 0.2, 0.6 and 1 samples at room-temperature. The (*hkl*) values corresponding to Bragg peaks are marked. (b) Variation of lattice constants with Si concentration. The error bars are within the symbols.

FIG. 2: (Color online) (a) Temperature dependence of magnetization for various compositions at 200 Oe applied field. Inset shows the magnetic transition region clearly. (b) $M$ vs $H$ curves over all the four quadrants for $x = 0$, 0.2, 0.6 and 1 samples at 5 K.

FIG. 3: Magnetization isotherms at various temperatures for the $x = 0$, 0.2, 0.6 and 1 samples.

FIG. 4: (Color online) Magnetic entropy change $-\Delta S_M$ vs temperature for $x = 0$, 0.2, 0.6 and 1 samples.

FIG. 5: $M^2$ vs $H/M$ isotherms at different temperatures close to the Curie temperature for $x = 0$, 0.2, 0.6 and 1 samples.

FIG. 6: (Color online) (a) $M$ vs $H$ on a log-log scale at $T_C$ for $x = 0$, 0.2, 0.6 and 1 samples. The solid line is the linear fit of Eq. 3. (b) Field dependence of the magnetic entropy change $-\Delta S_M$ for $x = 0$, 0.2, 0.6 and 1 samples. The solid lines are the fitted curves using Eq. 4.

FIG. 7: (Color online) Logarithmic scaling plot of $M|\varepsilon|^{-\beta}$ verses $H|\varepsilon|^{-(\beta+\gamma)}$ in the critical region. All experimental data fall on either of the two branches of the universal curve for $x = 0$, 0.2, 0.6 and 1 samples.

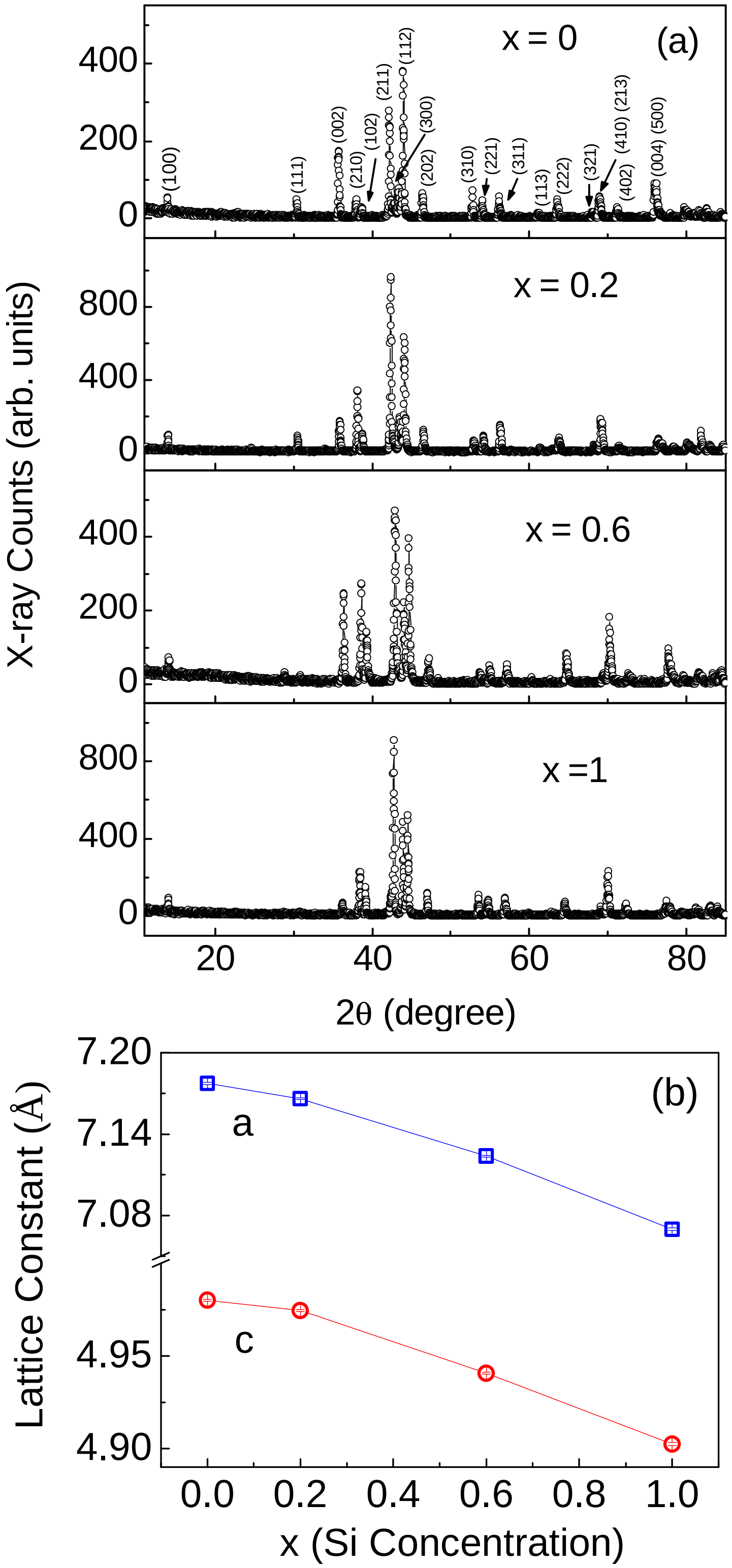

Fig. 1
Halder *et al*.

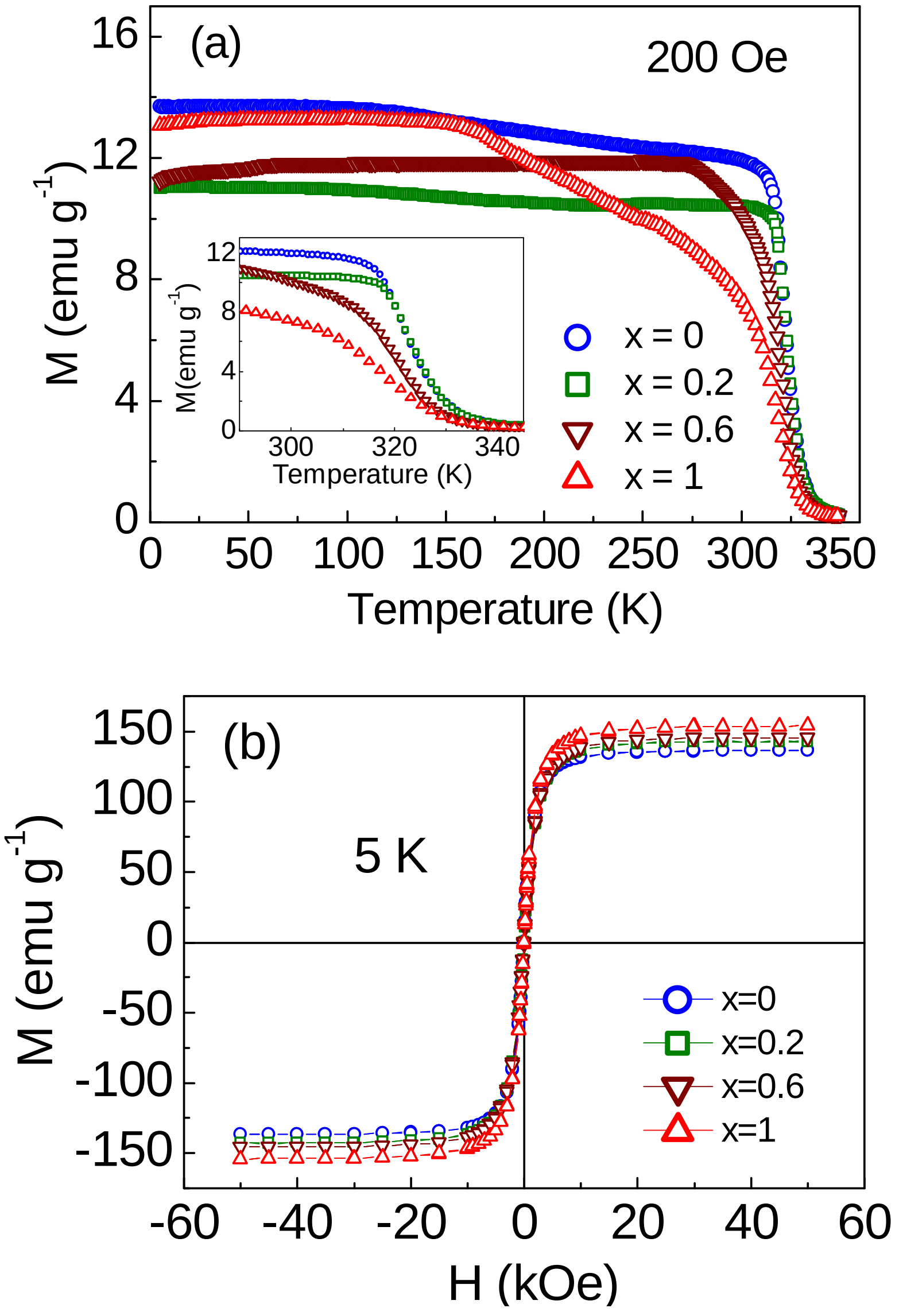

Fig. 2
Halder *et al*.

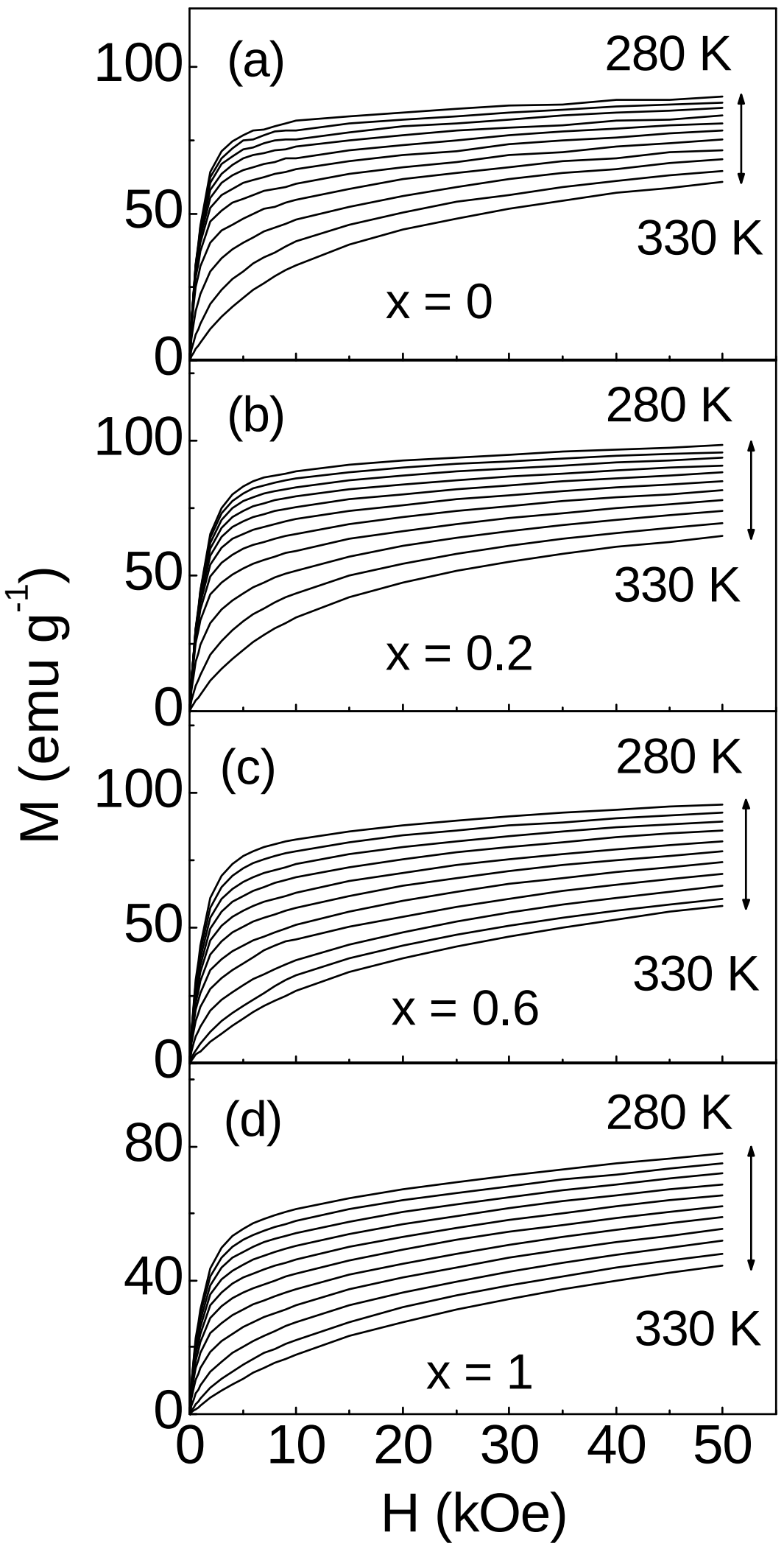

Fig. 3
Halder *et al*.

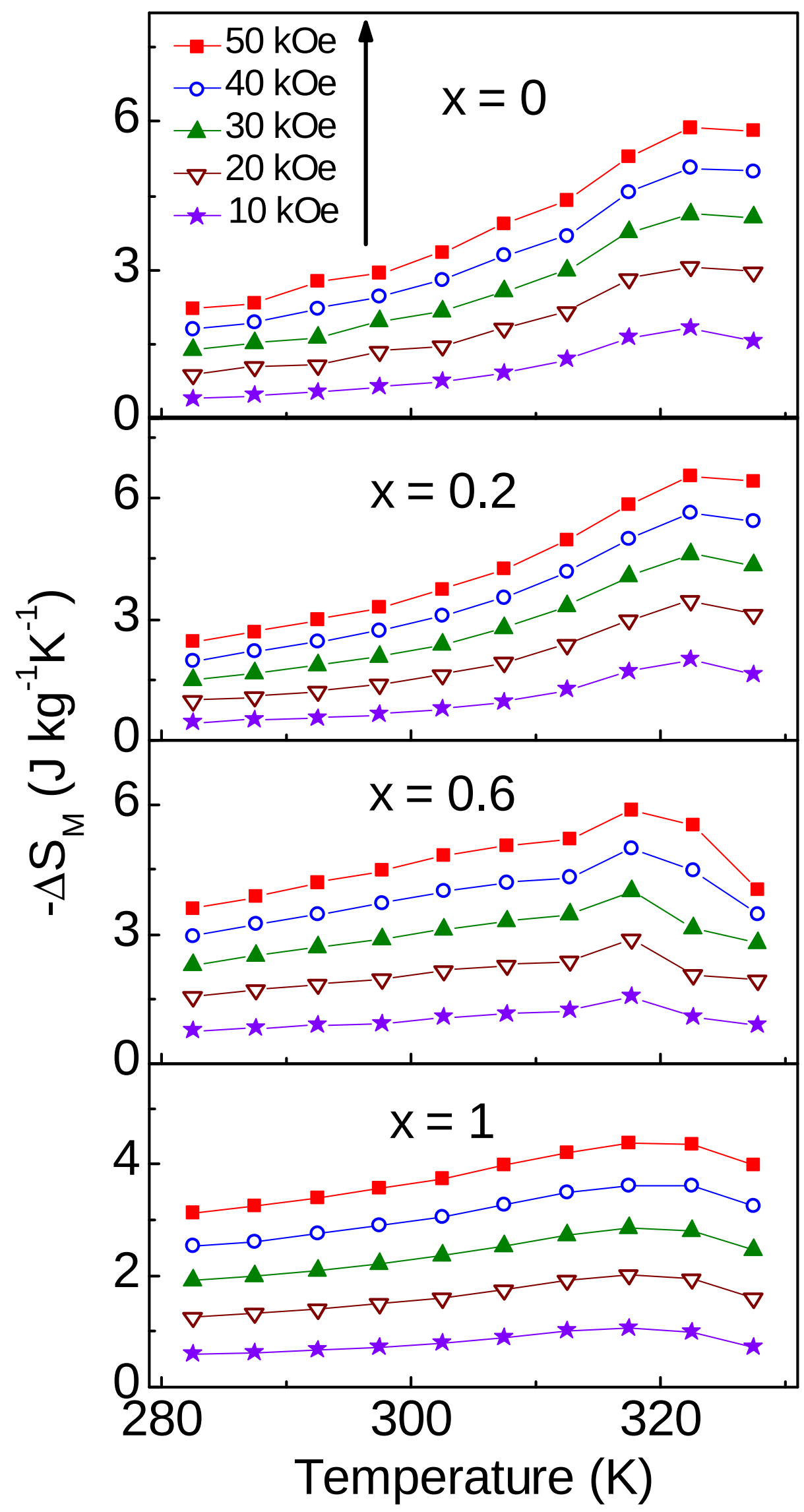

Fig. 4
Halder *et al*.

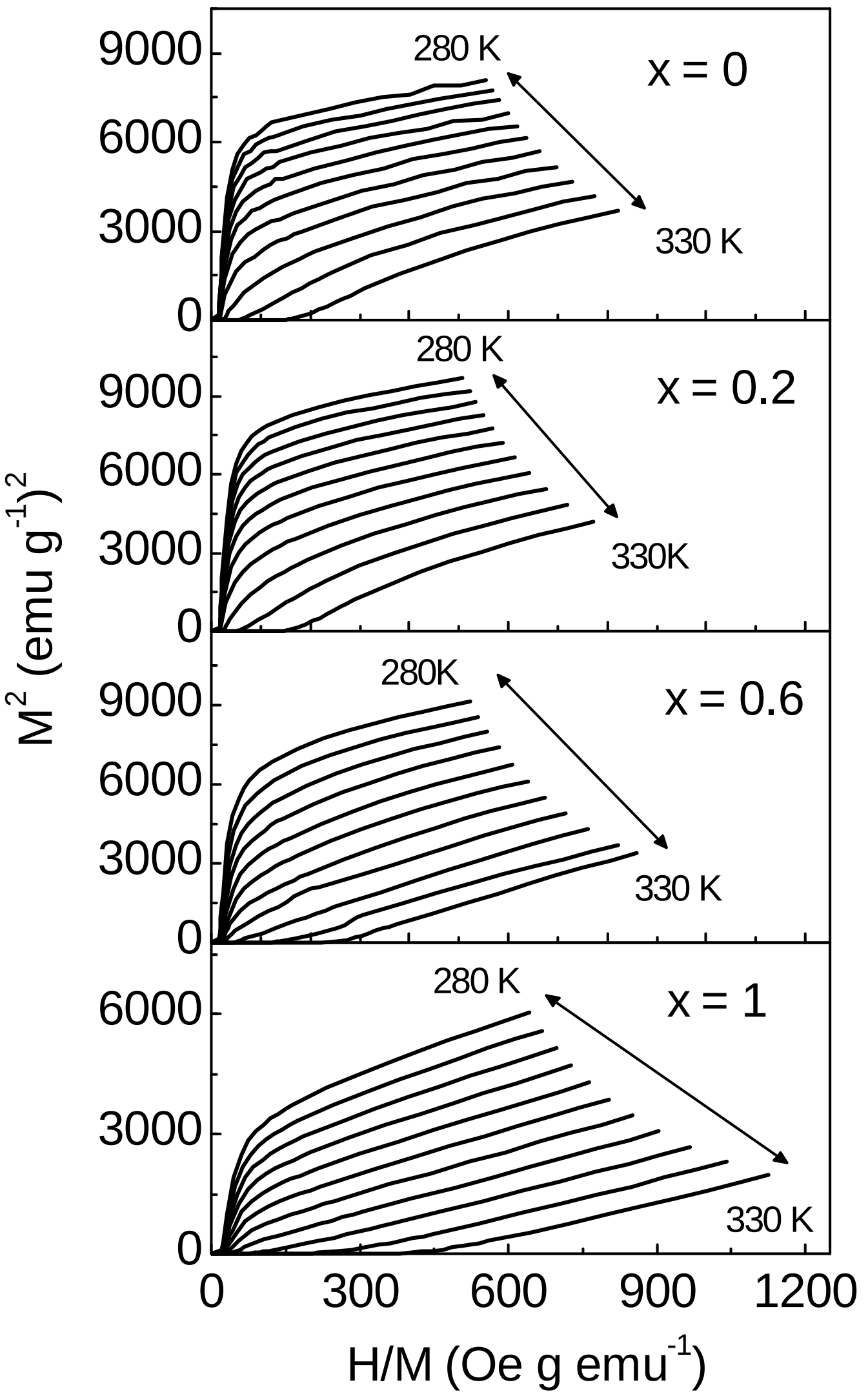

Fig. 5
Halder *et al*.

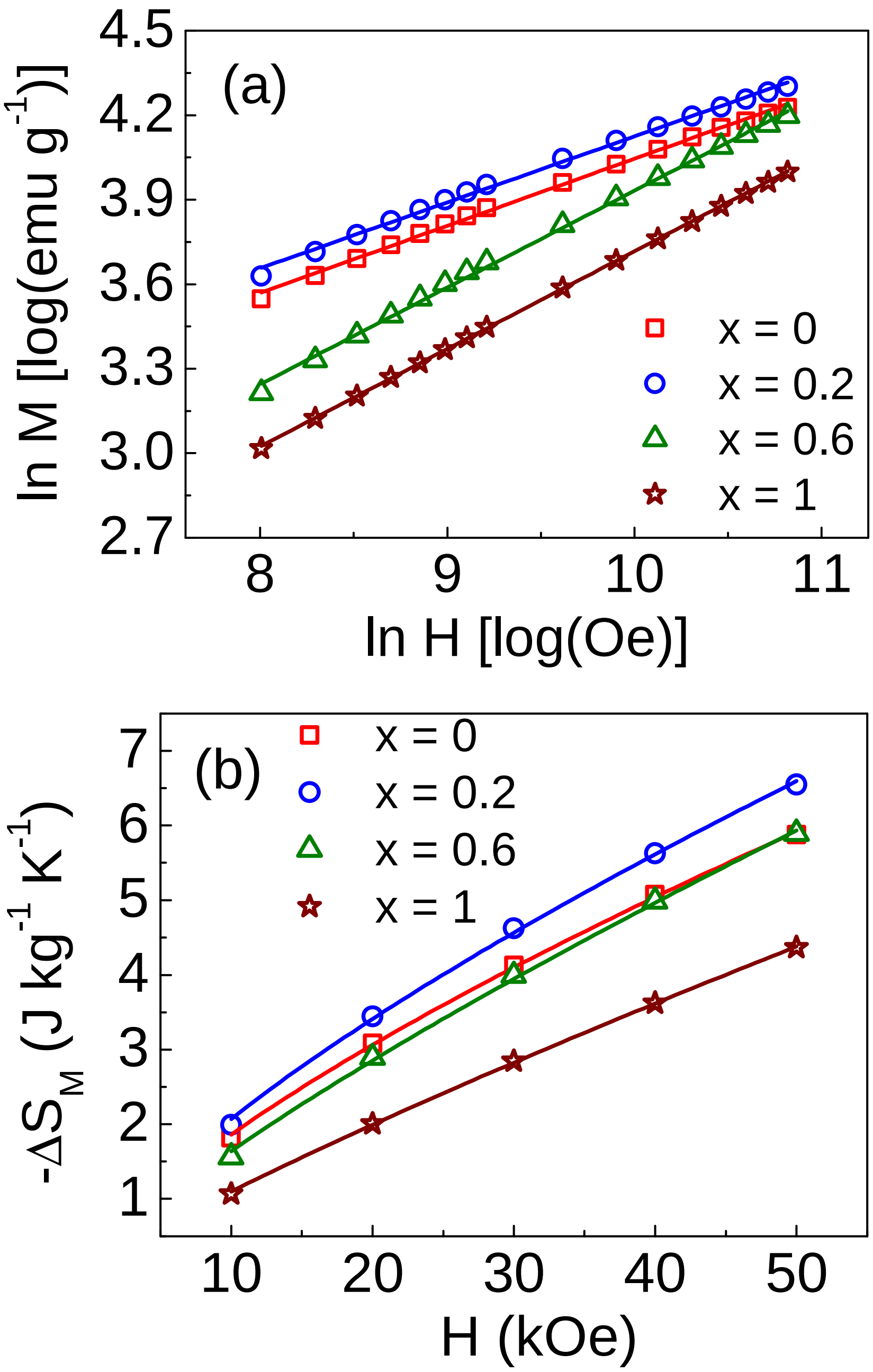

Fig. 6
Halder *et al*.

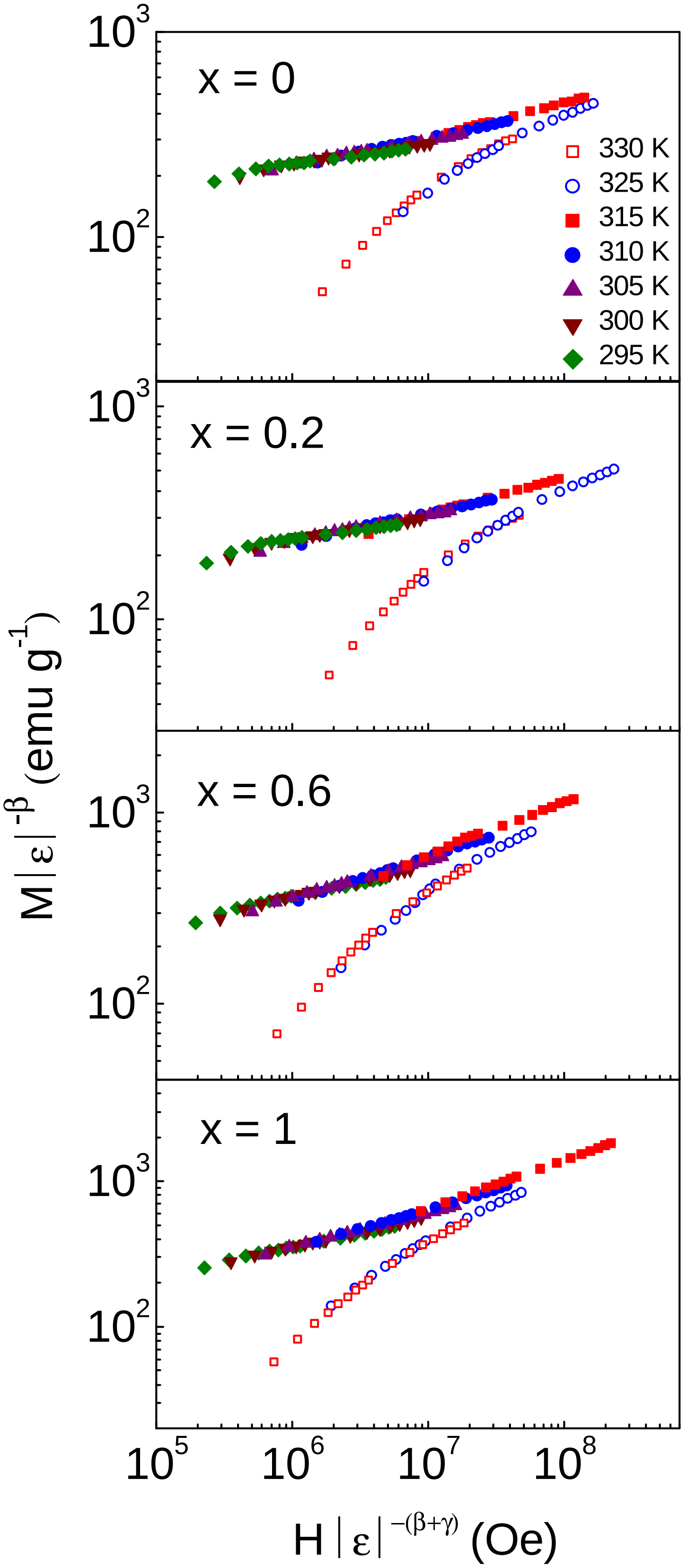

Fig. 7
Halder *et al*.